\documentclass[prb, twocolumn, nofootinbib]{revtex4-2}

\usepackage{bm}
\usepackage{graphicx}
\usepackage[x11names,dvipsnames]{xcolor}
\usepackage{amsmath}
\usepackage{amssymb}
\usepackage{amsthm}
\usepackage{bbold}
\usepackage{mathrsfs}
\usepackage{color}
\usepackage[unicode=true,colorlinks=true]{hyperref}
\usepackage[capitalize]{cleveref}

\usepackage{tikz}

\newcommand{\KP}{{\bf k}\ensuremath{\cdot}{\bf p}}
\newcommand{\qdn}{quantum dot (nanostructure)}
\newcommand{\qdns}{quantum dots (nanostructures)}

\newcommand{\bra}[1]{\left\langle {#1} \right\vert}
\newcommand{\ket}[1]{\left\vert #1 \right\rangle}
\newcommand{\brakt}[2]{\left\langle #1 \middle\vert #2 \right\rangle}
\newcommand{\braket}[3]{\left\langle #1 \left| #2 \right| #3 \right\rangle}

\newcommand{\rmi}{{\rm i}}
\newcommand{\rme}{{\rm e}}

\newcommand{\up}{{\uparrow}}
\newcommand{\down}{{\downarrow}}

\DeclareMathOperator{\Tr}{Tr}
\DeclareMathOperator{\diag}{diag}
\newcommand{\ebasis}{\bm{\mathcal{E}}}

\begin{document}

\title{Untangling the valley structure of states for intravalley \\ exchange anisotropy in lead chalcogenides quantum dots}
\author{I.D.~Avdeev}
\email{ivan.avdeev@mail.ioffe.ru}
\author{M.O.~Nestoklon}
%%\email{nestoklon@gmail.com}
\affiliation{Ioffe Institute, 194021 St. Petersburg, Russia}

\begin{abstract}
We put forward a generalized procedure which allows to restore the bulk-like electron and hole wave functions localized in certain valleys from the wave functions of quantum confined electron/hole states obtained in atomistic calculations of nanostructures.
As a demonstration, the procedure is applied to the lead chalcogenide quantum dots to extract the effective intravalley Hamiltonian of the exchange interaction for the ground exciton state PbS and PbSe quantum dots.
Renormalization of the anisotropic intravalley matrix elemets of velocity is also calculated.
The results demonstrate that the matrix elements of intravalley exchange in PbS quantum dots are much more anisotropic than ones in PbSe.
\end{abstract}

\maketitle

% ===============================================
\section{Introduction}
\label{sec:intro}

Currently, semiconductor-based nanostructures are widely used for various applications. In particular, quantum dots (QDs) \cite{Efros2021} offer the tunability of various properties from the basic ones like the effective band gap to more complicated such as exchange interaction in excitons\cite{Tamarat2023} and carrier $g$-factor values \cite{Avdeev2023g,Nestoklon2023}.
Rapid progress of experimental techniques demands for the detailed theoretical insight into properties of semiconductor nanostructures. However, until now there is a gap between purely phenomenological methods based on the \KP\ model \cite{Luttinger1955,Sercel1990} and atomistic calculations, both empirical \cite{Zunger_PP,Xavier_book} and \textit{ab initio} \cite{Hohenberg1964,Kohn1965}. For the band structure calculations, the interpretation of the atomistic results within the \KP\ framework is straightforward. For nanostructure calculations in most cases additional work has to be done. 
Observable quantities (splitting energies, optical matrix elements, etc.) are available directly from the atomistic calculations.
However, their values typically results from the complex interplay of conceptually different phenomena (mixing of the states at interfaces, anisotropy of effective masses, exchange interaction, etc.). For the qualitative description and prediction of physical properties of real nanosystems the values of interest are usually the latter ones.

Particularly complex problem is the fine energy structure of the nanostructures of multi-valley semiconductors (e.g. lead chalcogenides, Si, Ge), where such values of interest are blended by the mixing of the valley states.  
In simple cases, such as SiGe quantum wells \cite{Boykin2004,Nestoklon2006} or $[110]$-grown PbX nanowires with the simple surface \cite{Avdeev2019}, straightforward parametrization of the valley splitting is possible.
Though, in most cases the valley mixing can be taken into account only phenomenologically.
In this work we focus on lead chalcogenides and propose a generalized solution to this problem.

Lead chalcogenides Pb$X$, $X=$ S,Se are narrow direct band gap semiconductors suitable for infrared optoelectronic applications \cite{Sun2012,Sukhovatkin09,Tisdale2010} due to the tunability of the band gap in a wide range of the infrared spectrum.
Under normal conditions they have the rock-salt crystal structure with $O_h^5$ space group and complex multi-valley band structure. 
Conduction and valence band extrema in Pb$X$ are located at the four inequivalent anisotropic $L$ valleys.

As a result of the broken translational symmetry in lead chalcogenide \qdns different $L$ valley states are mixed and the carriers wave functions are the combinations of the pure valley states \cite{Poddubny2012,Avdeev2017,Avdeev2020}.
The valley mixing spreads the local density of states in k-space among all the $L$ valleys \cite{Avdeev2017} and makes it difficult to map the atomistic calculations onto the effective model and vice versa \cite{Avdeev2020}.
In particular, the very complex fine structure of the excitons stems from the interplay of spin-orbit splitting, valley-mixing and exchange interaction \cite{Avdeev2020}. This also results in complex behaviour of carriers' $g$-factors \cite{Avdeev2023g}. 
The second mechanism responsible for the lifting of the valley degeneracy is the anisotropy of effective masses.
The disentanglement of different contributions to the exciton fine structure is further complicated by the strong anisotropy of effective masses in PbX \cite{Svane10,Poddubny2012} which is, however, can be captured successfully in the framework of the effective \KP\ theory, see, e.g., Refs.~\cite{Bartnik10,Goupalov11,Goupalov2023}.

Below we present our solution of how to trace the valley structure of the ground electron (hole) states in PbX QDs from atomistic calculations to enable direct mapping onto analytical models.
The proposed procedure relies on the symmetry analysis and computation of the local density of states in reciprocal space at the four inequivalent $L$ points in the Brillouin zone.
As an example we use the obtained pure valley states to calculate the intravalley anisotropic exchange constants and intravalley interband velocity matrix elements in cubic, cuboctahedral and octahedral PbX quantum dots with tetragonal symmetry, similar to the ones studied in Ref.~\cite{Avdeev2020}.%, directly in the atomistic method.

We demonstrate that these quantities are almost insensitive to the shape of the quantum dot and agree very well with effective mass calculations.
Direct access to the valley states allows us to show that velocity and exchange Coulomb matrix elements in PbS quantum dots are much more anisotropic than in PbSe in the considered range of quantum dot diameters from 3 nm to 25 nm.

% ===============================================
\section{Valley states}
\label{sec:vstb}

In bulk PbX crystal the valley states are the electron and hole states at the band extrema located at the four inequivalent independent $L$ valleys.
In each valley the states are classified by irreducible representations of the $L$ valley wave vector point group $D_{3d}$ \cite{Avdeev2017}.
This group has six spinor representations: two two-dimensional $\Gamma_4^{\mp}$ and four one-dimensional conjugated $\Gamma_5^{\mp}$ and $\Gamma_6^{\mp}$ in Koster's notation \cite{Koster}.
These representations are also known as $L_6^{\mp}$, $L_4^{\mp}$ and $L_5^{\mp}$ respectively \cite{Dimmock64}.
(Hereafter throughout the paper we use only the Koster's notation.)
Since the one dimensional representations $\Gamma_5^{\mp}$ and $\Gamma_6^{\mp}$ are conjugated (related by time inversion) they also form doubly degenerate energy levels $\Gamma_5^{\mp}\oplus\Gamma_6^{\mp}\equiv\Gamma_{56}^{\mp}$.

\subsection{Extended effective mass model}

In PbX crystal the ground conduction band edge states are odd and transform according to $\Gamma_4^-$ ($L_6^-$), while the ground valence band edge states are even and form the basis of $\Gamma_4^+$ ($L_6^+$) \cite{Kang1997}.
The center of inversion is assumed to be at cation.
The standard basis functions of $\Gamma_4^{\pm}$ are (pseudo)spinors~\cite{Koster}, therefore we refer the band edge states at the $L$ valleys as the valley (pseudo)spinors
\begin{equation}
  \label{eq:vp}
  \ebasis_{\mu}^{b} = (\ket{b,\mu,\up},\ket{b,\mu,\down}),
\end{equation}
where $b=c,v$ or ``$-$'',``$+$'' is the band index, $\mu=0,1,2,3$ is the valley index and $\up,\down$ are the indices of (pseudo)spins oriented along the valley axis.
Due to the $O_h$ rotational symmetry of the bulk PbX crystal the valley (pseudo)spinors in different valleys are not fully independent.
Indeed, for any $g\in O_h$ the functions $g\ebasis_{\mu}^b \equiv \ebasis_{\mu'}^b$ are also eigenstates of the bulk Hamiltonian with same energy $E^b$, but some at different $L$ valley $g \bm k_{\mu}\equiv\bm k_{\mu}'$.
This allows us to construct the valley (pseudo)spinor basis
\begin{equation}
  \label{eq:vp4}
  \ebasis_{VP}^b = (\ebasis_{0}^b,\ebasis_1^b,\ebasis_2^b,\ebasis_3^b)
\end{equation}
of the irreducible star of $L$ valleys with known transformation properties 
\begin{equation}
  \label{eq:gvp4}
  g \ebasis_{VP}^b = \ebasis_{VP}^b D^b(g).
\end{equation}
We refer it as the ground valley multiplets.
The transformation matrices of $\ebasis_{VP}^b$ can be established explicitly by choosing specific rotations $g_{\mu} \in O_h$ to relate the valley states in different valleys, such as $\ebasis_{\mu}^b = g_{\mu} \ebasis_0^b$.
Possible choices for $g_{\mu}$ are either the powers of $S_{4z}$ rotoreflection \cite{Avdeev2020}, powers of $C_{4z}$ \cite{Goupalov2022} or $C_{2x},C_{2y}$ and $C_{2z}$ rotations \cite{Avdeev2023g}.

When nanostructured the translation symmetry is broken and the eightfold degenerate ground valley multiplets \eqref{eq:vp4} split into several energy levels. 
The number of levels and their symmetries are determined by decomposition of the transformation matrices $D^b(g)$, Eq.~\eqref{eq:gvp4}, into irreducible representations of the symmetry group of the \qdn.
Decomposition of the $D^b$ matrices is given by symmetrization matrices $S^b$ via ${S^b}^{-1}D^bS^b$.
The $S^b$ matrices are chosen in such a way so the new basis $\ebasis_{VP}^b S^b = \ebasis_P$
\begin{equation}
  \label{eq:Fz_mueta}
  \sum_{\mu=0}^4\sum_{\eta=\up,\down} \ket{b,\mu,\eta}S^b_{\mu\eta,\Gamma_i F_z} = \ket{b,\Gamma_i,F_z}
\end{equation}
transforms as (pseudo)spin.
Here $\ket{b,\Gamma_i,F_z}$ are the states which transform as the standard basis functions~\cite{Koster} of the irreducible representation $\Gamma_i$.
Several $S$ matrices for $[111]$-nanowires with $D_{3d}$ point group and QDs with $T_d$ symmetry were calculated in Ref.~\cite{Avdeev2017,Avdeev2020,Avdeev2023g}.
In \qdns\ the symmetry of the ground valley multiplets holds.
Indeed, the bulk states may be related to the states in nanostructures in the two-step procedure: (i) formation of combinations of bulk states into states which transform under representations of nanostructure symmetry group and (ii) renormalization of energies of these states due quantum confinement (localization in $r$-space and delocalization in $k$-space, see \cite{Luttinger1955}).

\subsection{Empirical tight-binding method}

For atomistic calculations of PbX \qdns\ we use the $sp^3d^5s^*$ nearest neighbour variant of the tight-binding method~\cite{Poddubny2012}.
In this model the electron and hole wave functions are expanded over the basis of L\"owdin orbitals $\ket{n\xi}$ \cite{Lowdin50} localized near atomic sites
\begin{equation}
  \label{eq:psi_tb}
  \ket{\Psi} = \sum_{n\xi} C_{n\xi} \ket{n\xi}.
\end{equation}
Here $\xi$ describes both spin and one of the $s,p,d$ or $s^*$ type of the orbital, $n$ is the atomic site index.
Coefficients $C_{n\xi}$ are obtained via numerical diagonalization of the tight-binding Hamiltonian, which is represented by a large sparse matrix 
\begin{equation}
  \label{eq:Htb}
  \hat H = \sum\limits_{n\xi n'\xi'} H_{n'\xi'n\xi} \ket{n'\xi'}\bra{n\xi}.
\end{equation}
In the nearest neighbour approximation there are up to $7\cdot2^2\cdot10^2\cdot N_a$ nonzero elements in the matrix, where $7=6+1$ is the maximal number of nearest neighbours $n\ne n'$ plus diagonal $n=n'$, $2$ is the number of spins, $10$ is the number of the $sp^3d^5s^*$ orbitals and $N_a$ is the number of atoms in the \qdn.

Numerical diagonalization of the tight-binding Hamiltonian is performed using the thick-restart Lanczos algorithm \cite{trlan2010, Nestoklon2016_VCA}.
This is efficient iterative method which has linear computational complexity $\mathcal O(m\cdot N_a)$ of finding exactly $m$ eigenvectors near the band gap.
Due to the Kramers symmetry the electron (hole) energy levels $E_i$ are doubly or quadruple degenerate depending on the point symmetry of the \qdn.
There is some randomness in degenerate tight-binding eigenstates $\ket{i,p}$ 
($p=1,\ldots,n_i$) of Eq.~\eqref{eq:Htb}
\begin{equation}
  \hat H \ket{i,p} = E_{i} \ket{i,p}
\end{equation}
since any linear combination of degenerate states is also an eigenstate with the same energy $E_{i}$.
For each energy level $E_{i}$ these states $\ebasis_i=(\ket{i,1},\ldots,\ket{i,n_i})$ form a basis of an irreducible representation $\Gamma_i$ of size $n_i$.
Therefore we can define the \textit{symmetrized} states $\ket{\Gamma_i,F_z}$ as
\begin{equation}
  \label{eq:psi_Fz}
  \sum_{p} \ket{i,p} V^{i}_{p,F_z} = \ket{\Gamma_i,F_z},
\end{equation}
where $V^{i}$ is a unitary matrix and $\ket{\Gamma_i,F_z}$ are
the (pseudo)spin-like states with standard transformation properties \cite{Koster}.
Eq.~\eqref{eq:psi_Fz} is readily generalized to any number of energy levels
\begin{equation}
  \label{eq:ep}
  \ebasis_{TB} V = \ebasis_P,
\end{equation}
where $\ebasis_{TB}=(\ebasis_1,\ldots,\ebasis_n)$ is the set of sets of tight-binding states with energies $E_{1},\ldots,E_{n}$, $\ebasis_P=(\ebasis_{\Gamma_1},\ldots,\ebasis_{\Gamma_n})$ is the set of sets of symmetrized states with standard transformation properties.
The matrix $V=\diag(V_1,\ldots,V_n)$ is block-diagonal.
Accidental degeneracy of states $E_i=E_j, i\ne j$ is not considered as may be removed by small perturbation.

\subsection{Mapping tight-binding results to effective mass model}

The next step is to identify the states of the split conduction and valence band ground valley multiplets in atomistic calculations.
In most cases they are simply eight closes to the band gap states in each band.
However, a situation when the distance between quantum confined levels is smaller than the splittings of the valley multiples is possible~\cite{Avdeev2017}.
For such extreme case the states of the ground valley multiplets can be identified in k-space.
As shown in Ref.~\cite{Avdeev2017} the ground electron and hole states have the maximum of local density in k-space exactly at the $L$ points of the Brillouin zone, while the maximum of excited states is slightly displaced.
In centrosymmetric \qdns\ excited states can be also distinguished by their opposite parity.

Finally, we define the conduction (valence) band valley states in PbX \qdns\ by combining Eqs.~\eqref{eq:Fz_mueta} and \eqref{eq:psi_Fz}.
In the matrix form it reads as
\begin{equation}
  \label{eq:EVP}
  \ebasis_{VP}^bS^b=\ebasis_{TB}^bW^b,
\end{equation}
where $\ebasis_{TB}^{c(v)}$ are the raw conduction (valence) band tight-binding states and $\ebasis_{VP}^b$ are the corresponding valley multiplets similar to Eq.~\eqref{eq:vp4}.
Instead of $V^b$, Eq.~\eqref{eq:psi_Fz}, we introduce a matrix $W^b$ to account for possible uncertainties of the raw states to enable the inverse transformation to the basis of valley (pseudo)spinors given by
\begin{equation}
  \label{eq:U}
  U^b=W^b{S^b}^{-1}.
\end{equation}
We factorize the matrix $W$ (the band index is omitted for brevity) into three matrices
\begin{equation}
  \label{eq:W}
  W = V P R\,,
\end{equation}
where $V=\diag(V_{i_1},\ldots,V_{i_N})$, Eq.~\eqref{eq:psi_Fz}, brings the tight-binding states to the (pseudo)spin form, the matrix $P$ arranges the order of irreducible representations and accounts for the phases of their bases and the matrix $R$ describes the possible rotation between repetitive irreducible representations in the decomposition of the valley multiplet.
The matrix $V$ can be computed as a sum over the point group of the \qdn\ as described in Appendix~\ref{app:sum_group}.
%The matrix $P = A\diag(p_{i_1}\mathbb1,\ldots,p_{i_N}\mathbb1)$ consists of the arranging permutation $A$ and phase multipliers $p_i$ for each irreducible representation $\Gamma_i$ in the decomposition.
%The unknown phases $p_i$ arise due to the uncertainty of the $V^i$ matrices, see details in Appendix~\ref{app:sum_group}.
%The matrix $R$ is an orthogonal matrix needed to account for possible interchange of bases of repetitive irreducible representations in the decomposition of the valley multiplet.
%This matrix is required, e.g., for [111]-nanowires with $D_{3d}$ symmetry, where the conduction and valence band ground valley multiplets split into $3\Gamma_4^{\mp}\oplus\Gamma_{56}^{\mp}$ levels~\cite{Avdeev2017}.
The extra matrix $P$ traces the permutations and phase multipliers for the irreducible representations in the decomposition, see details in Appendix~\ref{app:sum_group}.
Parametrization of the $R$ matrix is given in Appendix~\ref{app:irr_rot}.
The unknown phases and rotation angles for $P$ and $R$ matrices for the ground valley multiplets can be obtained by the numerical maximization of the local density of states in k-space at the four $L$ points of the Brillouin zone.
%The goal is to simultaneously maximize the local density of $\ket{c(v),\mu,\eta}$ states at $L_{\mu}$ valley and minimize it at others $L_{\mu'}, \mu'\ne\mu$.
%The optimization is fast since each iteration requires only several multiplications of $4\times8\times40$ Fourier images of states of the ground valley multiplets ($\times4$ valleys, $\times8$ states per multiplet and $20$ orbitals per Pb and X atoms) and $8\times8$ matrices $W S^{-1}$.
%Details of the Fourier transformation and computation of the local density of states in k-space are given in Ref.~\cite{Avdeev2017}.
%This procedure is numerically robust and can be applied to any nanostructure of any multivalley semiconductor.

\begin{figure}[tb]
\includegraphics[width=0.8\linewidth]{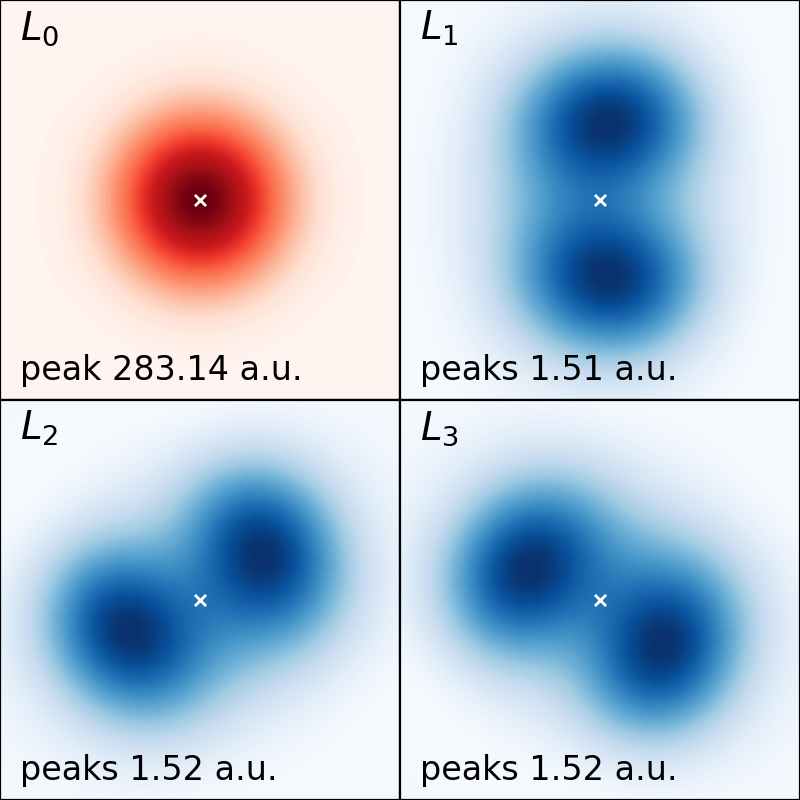}
\caption{Local density in the k-space of the same $\ket{c,0,\uparrow}$ valley state in the Brillouin Zone surface near $L$ valleys. The exact positions of the $L_{\mu}$ valleys are indicated by white ``x''. The main peak near $L_0$ valley (upper left) is shown in red colours, while the peaks near the other valleys $L_1, L_2, L_3$ are shown by blue colours with different scale of the colormap. The values (arb. units) of the maximal peak near each valleys are also indicated on the plots.}
\label{fig:kDOS_2D}
\end{figure}

\begin{figure}%[tb]
\includegraphics[width=1\linewidth]{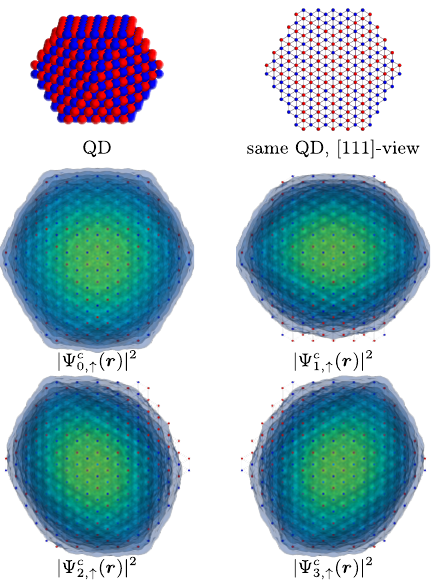}
\caption{$[111]$-view of the local densities in r-space of conduction band valley states $\ket{c,\mu,\up}, \mu=0,1,2,3$ in the cuboctahedral QD shown in inset. The size of atoms in the $[111]$-views are reduced to show the projection of contours of the wave functions $|\Psi_{\mu,\up}^c(\bm r)|^2$.}
\label{fig:rDOS_3D}
\end{figure}

To demonstrate the procedure, we reconstruct the valley states in the small cuboctahedral PbS QD without inversion ($D\approx3.2$ nm, $N=4,M=0$ in Ref.~\cite{Avdeev2020}). The arrangement of atoms in this QD is shown in the inset of Fig.~\ref{fig:rDOS_3D}.
In tetragonal quantum dots the ground conduction and valence band valley multiplets split into two doublets $\Gamma_6\oplus\Gamma_7$ and a quadruplet $\Gamma_8$.
%Therefore there is no need for the rotation matrix $R$ in the decomposition of $W$, Eq.~\eqref{eq:W}, since there is no repetitive irreducible representations in the decomposition.
The only fitting parameters are three discrete phase multipliers for each band.
%The $S^{c(v)}$ matrices, Eq.~\eqref{eq:Fz_mueta}, were calculated in Ref.~\cite{Avdeev2020} (via $S_{4z}$ rotations) and in Ref.~\cite{Avdeev2023g} (via $C_2$ rotations).
As a result we obtain eight states $\ket{b,\mu,\eta=\up,\down}$ in each band $b=c(v)$ localized near $L_{\mu}$ valleys.
The wave vectors of the $L$ valleys are chosen as
\begin{equation}
  \label{eq:kL}
  \bm k_0 \parallel [111],\quad
  \bm k_1 \parallel [\bar1\bar11],\quad
  \bm k_2 \parallel [1\bar1\bar1],\quad
  \bm k_3 \parallel [\bar11\bar1].
\end{equation}
The 2D local density of the $\ket{c,0,\up}$ state in k-space near the $L$ valleys is shown in Fig.~\ref{fig:kDOS_2D}.
The part of the density near the $L_0$ valley is shown by red, while the density near other valleys are shown by blue with different scale.
The scales of the colormaps are chose from 0 to the maximum of the local density in the corresponding cross section, which are indicated on the plots in arbitrary units.
One can see that the main peak at the $L_0$ valley is about $200$ times larger than the peaks near other $L_{\mu}, \mu=1,2,3$ valleys.
Moreover, these 2D plots reveal the admixture of excited states at different valleys since the maxima of their density are misplaced from the $L$ valleys.
The positions of the $L$ valleys in Fig.~\ref{fig:kDOS_2D} are shown by small white ``x'' at the middle of each plot.

In Fig.~\ref{fig:rDOS_3D} we also show the local density of all the valley states $\ket{c,\mu,\up}$, $\mu=0,1,2,3$ in r-space.
To emphasize the difference of the local densities of the valley states they are shown as projections onto the plane perpendicular to the $[111]$ axis.
In this plane projections of the effective masses in $L_0$ valley are isotropic, while projections of the other valleys $L_{\mu}$, $\mu\ne0$, are not.
This results in the anisotropy in the projections of the local densities of the $L_{\mu}$, $\mu=1,2,3$, valley states, which is clearly seen in Fig.~\ref{fig:rDOS_3D}.
The axis of this anisotropy depends on the projection of the corresponding $L$ valley onto the $(111)$ plane.
The state at the $L_0$ valley is isotropic in this projection.
The density of spin-down and valence band states look very similar.

% =======================================
\section{Results}

We apply the developed method to unwind the valley structure of states in cubic, cuboctahedral and octahedral PbS and PbSe quantum dots with tetragonal symmetry $T_d$, similar to the ones studied in Ref.~\cite{Avdeev2020}.
We show how having the explicit form of the valley multiplets helps to calculate physical properties of these quantum dots, such as interband velocity matrix elements and intravalley long range exchange Coulomb interaction, which otherwise are not directly accessible.

% -------------------------------------------
\subsection{Interband matrix elements of velocity}
\label{sec:VME}

\begin{figure}[tb]
\includegraphics[width=0.9\linewidth]{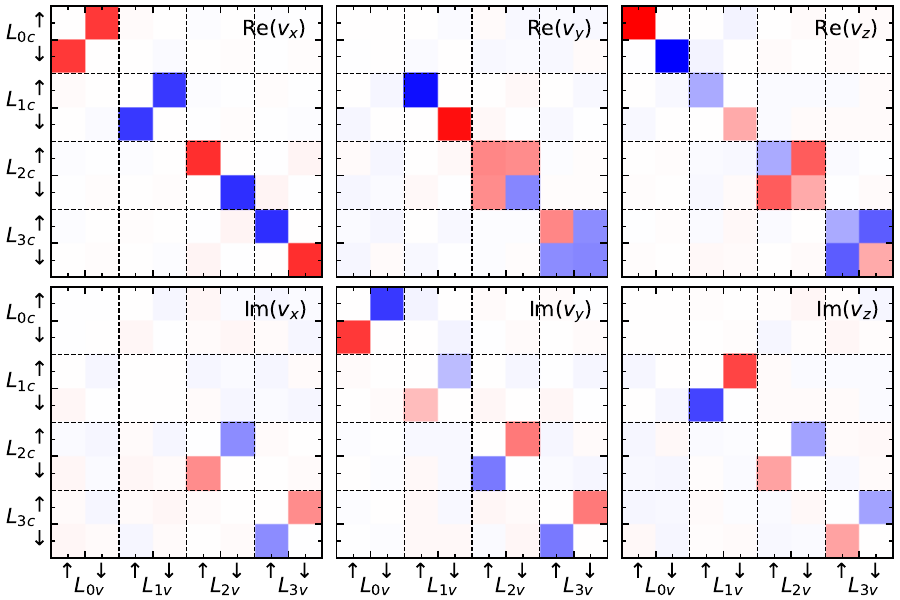}
\caption{Interband velocity matrix elements in the basis of valley state $\braket{c,\mu,\eta}{\hat{\bf v}}{v,\mu',\eta'}$ in the coordinate frame of the $L_0$ valley, Eq.~\eqref{eq:cf0}, in QD with $D\approx 3.2$ nm (see text). Color encodes the amplitude of corresponding matrix elements.}
\label{fig:VME_PbS}
\end{figure}

First we consider one of the simplest yet useful property of PbX quantum dots --- interband velocity matrix elements $\braket{c}{\hat{\bf v}}{v}$ between the electron and hole ground levels.
Velocity matrix elements in these quantum dots are computed as a commutator of of the tight-binding Hamiltonian with the coordinate operator, $\hat{\bf v} = i[\hat{H},{\bf r}]/\hbar$ and we assume the diagonal approximation ${\bf r}=\delta_{mn}\delta_{\alpha\beta}{\bf r}_n$ where ${\bf r}_n$ is the coordinate of $n$-th atom \cite{Ivchenko2002}.
Since the transformation of the tight-binding states to the valley (pseudo)spinors~\eqref{eq:EVP} given by the matrix $U^b$, $b=c,b$~\eqref{eq:U}, then the corresponding transformation of the velocity matrix elements is given by
\begin{equation}
  \bm V_{VP} = {U^c}^{\dag} \bm V_{TB} U^v.
\end{equation}
Result of this transformation is shown in Fig.~\ref{fig:VME_PbS} for the same $D\approx3.2$ nm cuboctahedral PbS quantum dot, which atomistic structure is shown in inset of Fig.~\ref{fig:rDOS_3D}.
In Fig.~\ref{fig:VME_PbS} the interband velocity matrix elements are given in the valley (pseudo)spin basis in the coordinates frame
\begin{equation}
  \label{eq:cf0}
  n_{0x}\parallel[1\bar10],\quad
  n_{0y}\parallel[11\bar2],\quad
  n_{0z}\parallel[111]
\end{equation}
of the $L_0$ valley.
Upper panels show real part of $v_x, v_y, v_z$ (from left to right) components of the velocity matrix elements, lower panels show imaginary parts.
The absolute values of velocity matrix elements are indicated by colour (positive by red, negative by blue, zero by white), the scale of the colormap is same for all the subplots.
The upper left corner in each subplot correspond to the intravalley velocity matrix elements in the $L_0$ valley.
One can clearly see that the spin matrices of the velocity operator in the $L_0$ valley are proportional to the Pauli matrices, $\hat v_i \propto \sigma_i$.  
The absolute values of the velocity matrix elements in $L_0$ valley, $v_x=v_y\ne v_z$, represent the internal anisotropy of the valley. 
One can also see there are traces of non-diagonal by the valley index optical transitions.
The intervalley interband velocity matrix elements are non-zero due to the admixture to the ground valley states of excited states in other valleys seen in Fig.~\ref{fig:kDOS_2D}.
This is a second order perturbation with respect to the valley splitting and it vanishes very quickly with increase of the size of the \qdn.

\begin{figure}[tb]
\includegraphics[width=0.9\linewidth]{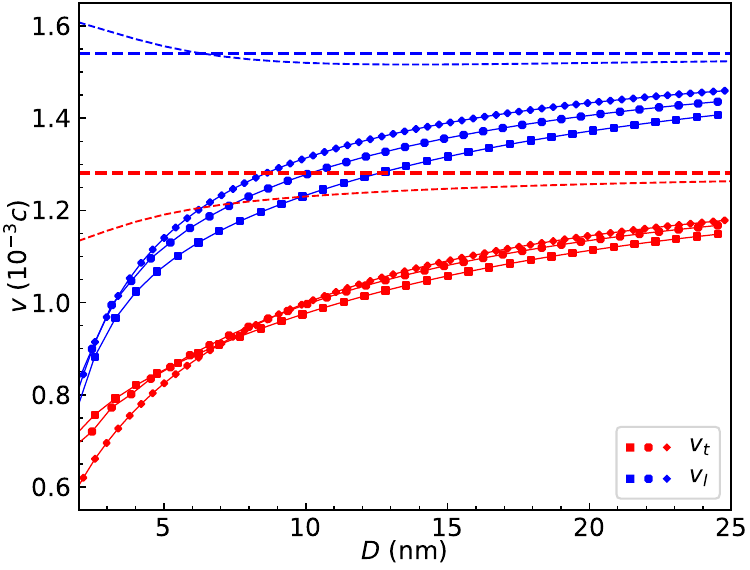}
\caption{Intravalley interband velocity matrix elements $v_t, v_l$ in $10^{-3}$ speed of light in PbS QDs with different shapes as a function of effective QD diameter. Dashed lines show corresponding interband velocity matrix elements in bulk crystal.}
\label{fig:VVV_PbS}
\end{figure}

\begin{figure}[tb]
\includegraphics[width=0.9\linewidth]{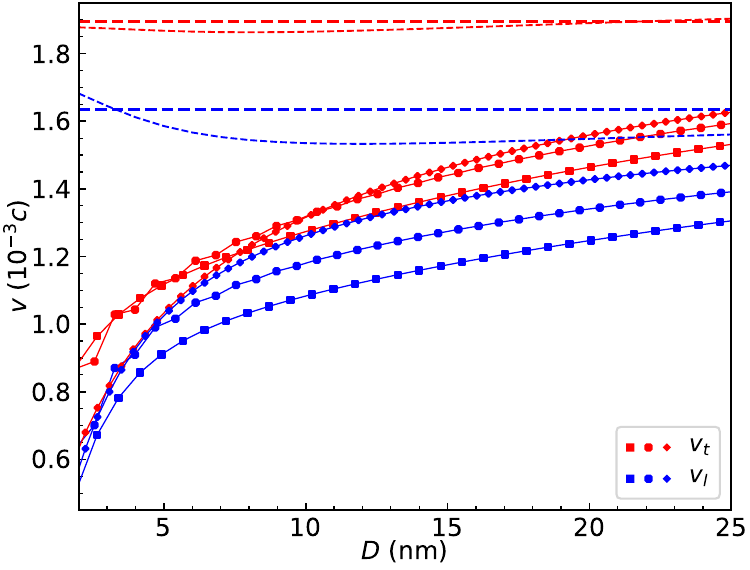}
\caption{Same as in Fig.~\ref{fig:VVV_PbSe} for PbSe QDs.}
\label{fig:VVV_PbSe}
\end{figure}

The structure of the velocity matrix elements in the basis of valley states, Fig.~\ref{fig:VME_PbS}, allows us directly calculate their values as
\begin{equation}
  \begin{array}{l}
    v_x = \operatorname{Re}\braket{c,0,\down}{\hat{v_x}}{v,0,\up}, \\
    v_y = \operatorname{Im}\braket{c,0,\down}{\hat{v_y}}{v,0,\up}, \\
    v_z = \operatorname{Re}\braket{c,0,\up}{\hat{v_z}}{v,0,\up}.
  \end{array}
\end{equation}
Results of the calculations are shown in Fig.~\ref{fig:VVV_PbS} for PbS and in Fig.~\ref{fig:VVV_PbSe} for PbSe quantum dots in thousandths of the speed of light.
Longitudinal ($v_l\equiv v_z$) intravalley interband velocity matrix elements are shown by blue, transverse ones ($v_t \equiv v_x=v_y$) by red.
Tight-binding data are shown by symbols connected by thin solid lines for each of the three considered shapes of the quantum dots: cubic by ``\tikz{\draw[x=1pt,y=1pt] (0,0)--(0,4)--(4,4)--(4,0)--(0,0);}'', cuboctahedral by ``\tikz{\draw[x=1pt,y=1pt] (1,0)--(3,0)--(4,1)--(4,3)--(3,4)--(1,4)--(0,3)--(0,1)--(1,0);}'' and octahedral by ``\tikz{\draw[x=1pt,y=1pt] (2,0)--(4,2)--(2,4)--(0,2)--(2,0);}''.
For the detailed description of the atomistic structure and shapes of the quantum dots see Ref.~\cite{Avdeev2020}.
Solid dashed lines show the corresponding velocity matrix elements in bulk PbS and PbSe crystals.
Data calculated within the framework of fully anisotropic \KP\ model~\cite{Avdeev2023g} are shown by thin dashed lines.

These plots, Figs.~\ref{fig:VVV_PbS} and~\ref{fig:VVV_PbSe}, reveal the scaling of the internal valley anisotropy in isotropic PbS and PbSe quantum dots.
One can clearly see that velocity matrix elements almost insensitive to the shape of the quantum dots and converge to the corresponding values in bulk crystal at large diameters.
Also these plots reveal that intravalley velocity matrix elements in PbSe quantum dots are almost isotropic in the range of diameters from about 5 nm to 15 nm, compared to to their PbS counterparts.
Another interesting result is that anisotropic \KP\ theory for spherical quantum dots fails to predict the scaling of velocity, even though it worked quite well the conduction and valence band confinement energy \cite{Avdeev2020}.

% -------------------------------------------
\subsection{Intravalley exchange anisotropy}
\label{sec:ALRX}

The second application of the valley pseuodspinors considered in this paper is the calculation of intravalley anisotropic long range exchange Coulomb interaction constants.
These constant are calculated here for the same PbS and PbSe quantum dots as in Sec.~\ref{sec:VME} and Ref.~\cite{Avdeev2020}.

The problem with calculation of intravalley exchange interaction constants is the complicated interplay of intra- and intervalley Coulomb interaction and the valley mixing of electron, hole and exciton states.
The full exciton Hamiltonian in PbX quantum dots has four main contributions
\begin{equation}
  \label{eq:HX}
  \hat H_X = \hat H_0 + \hat H_{VM} + \hat J + \hat K,
\end{equation}
where $\hat H_0 \equiv E_g(D)\mathbb1$ describes the quantum confinement of electrons and holes, $\hat H_{VM}$ is the valley mixing, $\hat J\equiv J\mathbb1$ is the direct Coulomb interaction and $\hat K$ is the long range exchange~\cite{Avdeev2020}.
The quantum confinement and direct Coulomb are trivial diagonal parts of the Hamiltonian.
The exciton fine structure is defined by the valley mixing $\hat H_{VM}$ and exchange interaction $\hat K$.
It was shown in Ref.~\cite{Avdeev2020} that intra- and intervalley exchange are equally important and a phenomenological model for the exchange matrix $\hat K$ was proposed.
The model is isotropic with one exchange constant $K(D)$ which can be calculated analytically.
In this model the electron-electron representation of the long range exchange Hamiltonian in one $L$ valley is
\begin{equation}
  \label{eq:H1Xiso}
  \hat{H}_{\text{exch}}^{\text{iso}} = K \left(\frac{\mathbb 1}2 - \frac{\bm \sigma^* \bm\sigma}{6}\right),
\end{equation}
where $K$ is the exchange constant and the spin matrix results from the angular parts of exchange integrals on the spherically symmetric electron and hole wave functions.
The model also takes into account the valley splittings as external parameters and allows to calculate absorption spectrum of the ground exciton level which is then can be compared to the similar spectra in tight-binding.
Comparison of the phenomenological model and tight-binding absorption spectra in PbS quantum dots was used in Refs.~\cite{Avdeev2020} and~\cite{Avdeev2022err} to estimate the exchange constant in the tight-binding.
The obvious drawback of this approach is that it requires a fitting procedure to minimize the difference between absorption spectra.
This approach is not accurate.

In this work we calculate the intravalley exchange interaction constants straightforwardly fully taking into account the valley anisotropy.
As discussed in Sec.~\ref{sec:vstb} the $L$ valley states, Eq.~\eqref{eq:vp}, transform according to $\Gamma_4^{-}$ and $\Gamma_4^+$ irreducible representation of the wave vector group $D_{3d}$ in conduction and valence bands, respectively.
Therefore in the direct product $\Gamma_4^-\otimes\Gamma_4^+ = \Gamma_1^-\oplus\Gamma_2^-\oplus\Gamma_3^-$ there are two one-dimensional, $\Gamma_1^-$, $\Gamma_2^-$, and one two-dimensional, $\Gamma_3^-$ irreducible representations, each corresponding to an exchange constant $K_s, K_l$ and $K_t$.
As a result, the electron-electron representation of the Hamiltonian of the anisotropic intravalley exchange interaction has the following form
\begin{equation}
  \label{eq:H1Xaniso}
  \hat{H}_{\text{exch}}^{\text{aniso}} =
  K_s \frac{\mathbb1}2 -
  K_t \frac{(\sigma_x^*\sigma_x + \sigma_y^*\sigma_y)}{6} -
  K_l \frac{(\sigma_z^*\sigma_z)}{6}.
\end{equation}
Here we used same spin matrices as in Eq.~\eqref{eq:H1Xiso}.
For the known intravalley exchange matrix the exchange constants $K_s$, $K_t$ and $K_l$ are calculated as
\begin{equation}
  K_s = \frac{\Tr(\hat H_{\text{exch}}^{\text{aniso}})}{2}
  \,,\quad
  K_{t(l)} =-\frac{3\Tr(M_{t(l)}^*\hat H_{\text{exch}}^{\text{aniso}})}{2}\,,
\end{equation}
where $M_t = \sigma_x^*\sigma_x$ or $\sigma_y^*\sigma_y$ and $M_l = \sigma_z^*\sigma_z$.
Since the energy of the dark excitons are not shifted by the exchange \cite{goupalov2003a}, it is natural to expect that
\begin{equation}
  \label{eq:Ks}
  K_s = \frac{2K_t+K_l}{3}
\end{equation}
and the eigenvalues of $\hat{H}_{\text{exch}}^{\text{aniso}}$ are
\begin{equation}
  \label{eq:Estl}
  E_s=0,\quad
  E_t=\frac{K_t+K_l}3,\quad
  E_l=\frac{2K_t}3\,.
\end{equation}
The intravalley part \eqref{eq:H1Xaniso} of the full $64\times64$ exhcange Hamiltonian can also be calculated via the inverse transformation~\eqref{eq:U}. 
For the $v\otimes c$ exciton indexing scheme the transformation of exciton states is given by the direct product ${U^v}^* \otimes U^c$.

\begin{figure}[tb]
\includegraphics[width=0.9\linewidth]{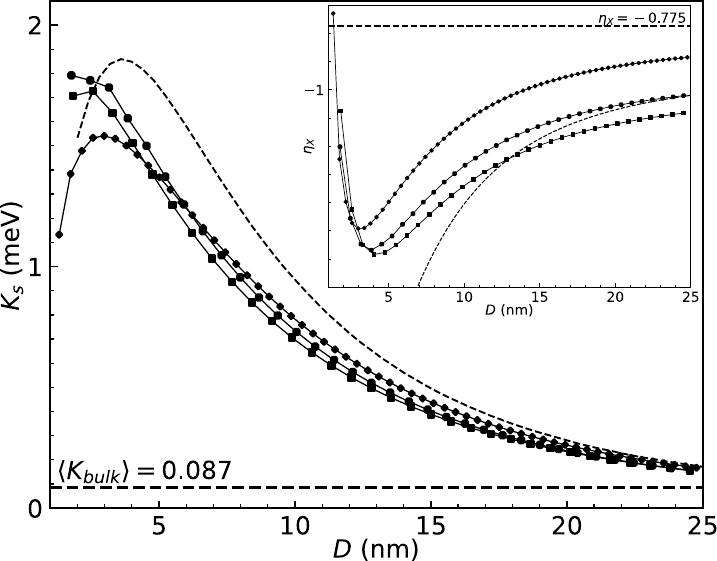}
\caption{Top panel shows the averaged intravalley exchange constant $K_s$ as a function of QD diameter. Bottom panel shows the intravalley exchange anisotropy parameter $\eta_X$ as a function of QD diameter. Tight-binding data are shown by symbols connected by thin solid lines. Each line and type of symbols correspond to different QD shape: cubic, cuboctahedral and octahedral. Thin dashed lines show the data calculated in the framework of anisotropic \KP\ theory. Thick dashed line in the bottom panel shows the values of $\eta_X$ in the bulk crystal (see Table~\ref{tb:LR_X}).}
\label{fig:Kan_PbS}
\end{figure}

\begin{figure}[tb]
\includegraphics[width=0.9\linewidth]{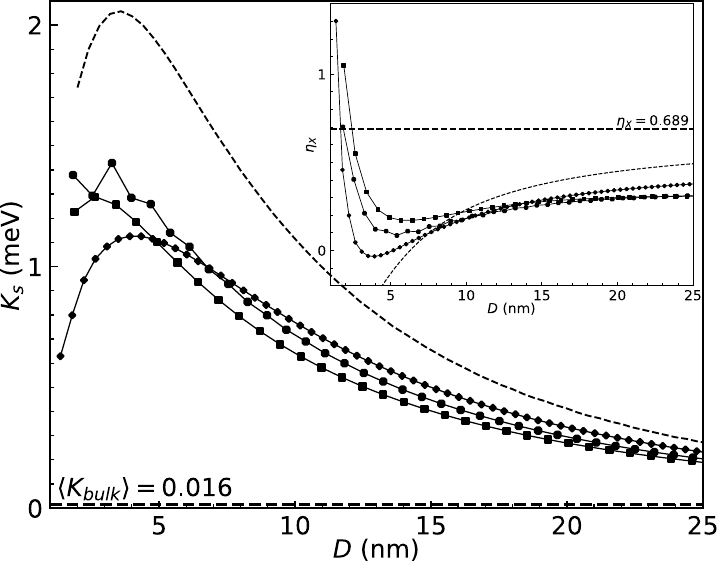}
\caption{Same as in Fig.~\ref{fig:Kan_PbS} but for PbSe QDs.}
\label{fig:Kan_PbSe}
\end{figure}

Results of the calculations are shown in Fig.~\ref{fig:Kan_PbS} for PbS and in Fig.~\ref{fig:Kan_PbSe} for PbSe cubic, cuboctahedral and octahedral QDs with tetragonal symmetry.
Instead of $K_s,K_t,K_l$ in Figs.~\ref{fig:Kan_PbS} and \ref{fig:Kan_PbSe} we show $K_s(D)$ and the long range exchange anisotropy parameter
\begin{equation}
  \eta_X(D) = \frac{K_l(D)-K_t(D)}{K_s(D)}.
\end{equation}
Tight-binding data are shown by symbols connected by thin solid lines for each of the three considered shapes of the QDs: cubic by ``\tikz{\draw[x=1pt,y=1pt] (0,0)--(0,4)--(4,4)--(4,0)--(0,0);}'', cuboctahedral by ``\tikz{\draw[x=1pt,y=1pt] (1,0)--(3,0)--(4,1)--(4,3)--(3,4)--(1,4)--(0,3)--(0,1)--(1,0);}'' and octahedral by ``\tikz{\draw[x=1pt,y=1pt] (2,0)--(4,2)--(2,4)--(0,2)--(2,0);}''.
Detailed description of the atomistic structure and shapes of the QDs are given in Ref.~\cite{Avdeev2020}.
Data calculated within the framework of fully anisotropic \KP\ model~\cite{Avdeev2023g} are shown by thin dashed lines.
One can see that the intravalley exchange is almost insensitive to the shape of the QDs and does not oscillate with the change of the QD diameter.
The tight-binding results agree well with anisotropic \KP\ theory.
Details of the \KP\ calculations are given in Appendix~\ref{app:kpX}.

The asymptotic values for $\eta_X$ may be calculated from \KP\ theory \cite{Gupalov1998,Gupalov2000}. 
The longitudinal-transverse splitting reads as 
\begin{equation}
 \hbar \omega_{LT} = \frac{8 e^2 \hbar^2 P^2}{\varepsilon_{\infty} m_0^2 E_g^2 a_B^3}
\end{equation}
where $a_B$ is the exciton Bohr radius. It may be shown that the effect of the valley anisotropy may be fully accounted by considering the direction-dependent longitudinal-transverse splitting 
\begin{equation}
 \hbar \omega_{LT}^{l(t)} = \frac{8 e^2 \hbar^2 P^2_{l(t)}}{\varepsilon_{\infty} m_0^2 E_g^2 a_B^3}\,.
\end{equation}
Then, the anisotropy of the main components of longitudinal-transverse splitting of the exciton is 
\begin{equation}
  \eta_X = \frac{\omega_{LT}^l-\omega_{LT}^t}{\left\langle \omega_{LT} \right\rangle} = 6\frac{P_t^2-P_l^2}{2P_t^2+P_l^2}.
\end{equation}
The values of $\eta_X$ for bulk PbS and PbSe are also given in Table~\ref{tb:LR_X}.

Similarly to the intravalley exchange, Eq.~\eqref{eq:H1Xaniso}, one can also compute the non-diagonal intervalley parts $\hat{H}_{\text{exch}}^{\mu\mu'}$, $\mu\ne\mu'$ of the full exchange Hamiltonian. 
The magnitude of the intervalley exchange in highly symmetric quantum dots is almost identical to the intravalley one $|\hat{H}_{\text{exch}}^{\mu\mu'}|\sim|\hat{H}_{\text{exch}}^{\mu\mu}|$, so the full exchange Hamiltonian is almost isotropic and also leads to the formation of the ultra-bright valley-symmetric superradiant exciton triplet as was shown in simplified model in Refs.~\cite{Avdeev2020} and \cite{Goupalov2022}.
However, the intervalley exchange has very complicated non-analytical form and we do not present it here.
The only difference with the isotropic approximation is that the valley anisotropy leads to the brightening of some dark triplets. Though, their oscillator strength is still two order of magnitude smaller and the use of isotropic model \cite{Avdeev2020} is fully justified.

\begin{table}
\caption{Main components of longitudinal-transverse long range exchange exciton splittings and material parameters for bulk PbS and PbSe. Momentum matrix elements are given in atomic units.}
% $\hbar=m_0=|e|=1$
\label{tb:LR_X}
\begin{ruledtabular}
\begin{tabular}{c|cc}
   & PbS & PbSe \\
   \hline
   $\hbar \omega_{LT}^t$ ($\mu$eV) & $76$ & $19$ \\
   $\hbar \omega_{LT}^l$ ($\mu$eV) & $110$ & $11$ \\
   $\left\langle \hbar \omega_{LT} \right\rangle$ ($\mu$eV) & $87$ & $16$ \\
   $\eta_X$ & $-0.775$ & $0.689$ \\ 
   \hline
   $P_t$ (atomic) & $0.1756$ & $0.3086$ \\
   $P_l$ (atomic) & $0.2110$ & $0.2293$ \\
   $E_g$ (eV) & $0.294$ & $0.213$ \\
   $\varepsilon_{\infty}$ \cite{Zemel1965} & $19.2$ & $26.9$ \\
   $a_B$ (nm) \cite{Moreels2009,Wise2000,Lifshitz2003} & 18 & 46 \\
\end{tabular}%
\end{ruledtabular}%
\end{table}

% ===============================================
\section{Conclusions}

To conclude, we developed a generalized procedure which allows us to restore the valley states in nanostructures of multivalley semiconductors starting from the quantum confined electron/hole states obtained in atomistic calculations.
The method allows for the extraction of the parameters of effective Hamiltonians and/or observables in the physically transparent basis and makes it possible for the direct mapping of the results of atomistic calculations to simplified analytical models.

To demonstrate the strength of the procedure we directly extract the anisotropic exchange constants in PbS and PbSe faceted (cubic, cuboctahedral and octahedral) QDs from the empirical tigh-binding calculations and compared the results to the fully anisotropic \KP\ model.
We also showed that in PbSe the intravalley exchange interaction, as well as the intravalley velocity, is almost isotropic, while in PbS it is strongly anisotropic with its longitudinal part being close to zero.

The developed procedure does not depend on particular symmetry, periodicity or valley composition of states of the considered nanostructure and therefore can be applied for any atomistic calculations (including density functional theory) of any multivalley semiconductor nanostructure.

% ===============================================
\section*{Acknowledgments}
Authors thank S.V~Goupalov, M.M~Glazov and E.L.~Ivchenko for fruitful discussions.
The work of IDA was supported by Russian Science Foundation under grant no. 22-72-00121.
IDA also thanks the Foundation for Advancement of Theoretical Physics and Mathematics ``BASIS''.
%, \url{https://rscf.ru/en/project/22-72-00121/}

% ===============================================
% ===============================================
\begin{appendix}

% ===============================================
\section{Symmetrization}
\label{app:sum_group}

Let $\ebasis=(\chi_1,\chi_2,\ldots,\chi_d)$ be the basis of an irreducible representation $\Gamma$ of a finite group $G$ of size $N$, which transforms $\forall g\in G$ as $g \chi_i = \chi_{j} T_{ji}(g)$ or $g \ebasis = \ebasis T(g)$ and $V$ be a unitary matrix.
Then $\ebasis'=\ebasis V$ also forms the basis of the same irreducible representation, but transforms $\forall g\in G$ as $g \ebasis' = \ebasis' D(g)$, where 
\begin{equation}
  \label{eq:V}
  \forall g\in G:~D(g) = V^{-1}T(g)V\,.
\end{equation}

{\it Lemma:} If $D(g)$ (desired) and $T(g)$ (target) matrices are known, then the matrix $V$ can be calculated as
\begin{equation}
  \label{eq:Vn}
  V_n = \alpha\,\rme^{\rmi \frac{2\pi n}{d}} \frac{\tilde V}{\det(\tilde V)^{\frac1d}}
  \,,\quad
  \tilde V = \sum_{g\in G} T(g) U D(g^{-1})\,.
\end{equation}
Here $d$ is the dimension of the representation $\Gamma$, $\alpha$ is a common phase multiplier, $|\alpha|=1$, and $U$ is a matrix with one or several ones.
Additional phase $\exp(\rmi2\pi n/d)$ and index $n=0,\ldots,d-1$ are added explicitly to underline the uncertainty of the root of the complex determinant.

{\it Proof.}
Consider the matrix $D(g)V^{-1}\tilde V$, $g\in G$.
Substituting $\tilde V$ from Eq.~\eqref{eq:Vn} we obtain
\begin{equation}
  \label{eq:DVV}
  D(g) V^{-1}\tilde V = D(g) \sum_{g'} V^{-1} T(g')UD(g'^{-1}).
\end{equation}
Now using Eq.~\eqref{eq:V} $D(g)V^{-1}T(g')=V^{-1}T(g)T(g')\equiv V^{-1}T(g'')$ and expanding $g'^{-1}$ as $(gg')^{-1}g\equiv g''^{-1}g$ we transform Eq.~\eqref{eq:DVV} to
\begin{equation}
  \label{eq:VVD}
  V^{-1}\sum_{g''} T(g'')UD(g''^{-1}) D(g)= V^{-1}\tilde V D(g).
\end{equation}
Following the Shur's lemma \cite{BirPikus} the matrix $V^{-1}\tilde V$ is proportional to the unit matrix and its trace equals to $\Tr(V^{-1}\tilde V) = N \Tr(V^{-1}U)$.
Since $V$ is unitary, then $V^{-1}$ has at least one nonzero element $V_{ik}^{-1}\ne 0$.
Let $U_{jl} = \delta_{jk}\delta_{il}$, then $\Tr(V^{-1}U) = \sum_{l,j} V_{lj}^{-1}U_{jl} = V_{ik}^{-1} \ne 0$. 
Therefore $\tilde V \propto V$, $\det \tilde V\ne 0$ and $V_n$ is unitary matrix satisfying Eq.~\eqref{eq:V}. \qedsymbol{}
Notice the proof is very similar to the proof of the great orthogonality theorem \cite{BirPikus,Zee2016}.

The unknown matrix $U$ can be found numerically simply by searching through all $d^2$ square matrices with one nonzero element.
The phase multiplier $\alpha$ is chosen to adjust the time inversion symmetry given by the complex conjugation operator $\hat K$.
Indeed, let $\hat K \ebasis = \ebasis K$, then $\hat K\ebasis'= \ebasis' V^{-1}KV^* \equiv \ebasis'K'$ and the matrix $K' \propto {\alpha^*}^2$.
The phase $\exp(\rmi 2\pi n/d)$ should also be adjusted.
The sum over the group elements \eqref{eq:Vn} assumes either sum over double group \cite{Koster} or use of projective representations \cite{BirPikus}. In calculations, it is more convenient to sum over all different matrices which can be obtained from the set of matrices of the group $G$ generators, which gives the same result.

Equation \eqref{eq:Vn} can be generalized for projective representations
\begin{equation}
  \label{eq:Vnproj}
  V_n \propto \tilde V\,,\quad\tilde V = \sum_{r\in F_{\bm k}} \frac{T(r) U D(r^{-1})}{\omega(r^{-1},r)},
\end{equation}
where $r$ are rotations of the point group $F$ and $\omega(r,r')$ is the factor system of the irreducible representation \cite{BirPikus}. 
Notice $\omega$ must be the same for $T$ and $D$ matrices, which can be achieved via bringing both $\omega_T$ and $\omega_D$ to the standard form.
The proof of Eq.~\eqref{eq:Vnproj} is similar, except the following property of the factor system should be used
\begin{equation}
  \omega\left((rr')^{-1},rr'\right) = \frac{\omega(r'^{-1},r')\omega((rr')^{-1},r)}{\omega(r,r')}\,.
\end{equation}
This is general property of the factor system \cite{BirPikus}
\begin{equation}
  \omega(h_1,h_2h_3)\omega(h_2,h_3)=\omega(h_1h_2,h_3)\omega(h_1,h_2)\,,
\end{equation}
where $h_1=(rr')^{-1}$, $h_2=r$ and $h_3=r'$.

Formulae Eqs.~\eqref{eq:Vn} and \eqref{eq:Vnproj} can be applied also for reducible representations such as $\Gamma_{i_1}\oplus\Gamma_{i_2}\ldots\oplus\Gamma_{i_N}$ with no repetitive irreducible ones in the sum.
Otherwise the Shur's lemma breaks and other techniques should be used.
However, this problem is not typical for tight-binding or other atomistic methods, since in most cases each energy level corresponds to single irreducible representation and accidental degeneracies are extremely rare.

% ===============================================
\section{Interchange of basis functions}
\label{app:irr_rot}

Let $\Gamma=\bigoplus_{i=1}^n\Gamma_m$ be a reducible representation of a group $G$ consisting of $n$ repetitive equivalent irreducible representations $\Gamma_m$ of size $m$.
Let
\begin{equation}
  \label{eq:Enm}
  \ebasis = (\underbrace{\ebasis_m^1,\ebasis_m^2,\ldots,\ebasis_m^n}_{n})\,,
  \quad
  i\ne j\implies \ebasis_m^i \ne \ebasis_m^j\,,
\end{equation}
be its basis which transforms under $g\in G$ as
\begin{equation}
  \label{eq:gEnm}
  g \ebasis = \ebasis D(g) 
  \equiv \ebasis \mathbb 1_n \otimes D_m(g)
\end{equation}
and has time reversal symmetry
\begin{equation}
  \label{eq:KEnm}
  \hat K \ebasis = \ebasis^* T
  \equiv \ebasis \mathbb 1_n \otimes T_m(g).
\end{equation}
Here $\mathbb 1_n$ is the $n\times n$ unit matrix.

The basis \eqref{eq:Enm} is not unique.
Indeed, consider an interchange of basis functions
\begin{equation}
  \ebasis'=\ebasis U
\end{equation}
given by a matrix $U$.
To satisfy Eqs.~\eqref{eq:gEnm} and \eqref{eq:KEnm} the new basis should be i) orthonormal $\ebasis'^{\dag}\ebasis' = \mathbb1$ and should transform under $g\in G$ and $\hat K$ by the same matrices ii) $g \ebasis' = \ebasis'D(g)$ and iii) $\ebasis'^* =\ebasis' T$.
These conditions constrain the matrix $U$ to:
i) $U^{\dag}U=\mathbb1$,
ii) $\forall g\in G: D(g)U=UD(g)$
and 
iii) $UT=TU^*$.
Due to the Schur's lemma \cite{BirPikus} and the structure of the transformation matrices $D(g)=\mathbb1_n\otimes D_m(g)$ the matrix $U$ matrix is a direct product $U_n \otimes \mathbb1_m$ of $n\times n$ matrix $U_n$ and the $m\times m$ unit matrix $\Bbb 1_m$.
Since $T=\Bbb 1_n\otimes T_m$ the third condition $UT=TU^*$ requires $U_n\in O(n)$ to be real orthogonal matrix.

For practical realization we construct the $U_n$ matrix as a product of $n(n-1)/2$ Givens rotation matrices $R_{ij}(\phi)$ \cite{Matteson2017} and a matrix of phases $P$ which sets the signs of the basis functions:
\begin{equation}
  \label{eq:UnRijIk_simple}
  U_n = P \prod_{i<j} R_{ij}(\tilde \phi_{ij})
  ~~\text{or}~~
  U_n = \left(\prod_{i<j} R_{ij}(\tilde \phi_{ij})\right)P
  \,,
\end{equation}
where $\tilde \phi_{ij} = \pm \phi_{ij}$ and
\begin{equation}
  P= \diag(1,p_2,\ldots,p_n)\,,\quad p_i = \pm 1\,.
\end{equation}
Nonzero matrix elements of $R_{ij}(\phi)$ are $[R_{ij}(\phi)]_{kk}=1, k\ne i,j$, $[R_{ij}(\phi)]_{ii}=[R_{ij}(\phi)]_{jj}=\cos(\phi)$ and $[R_{ij}(\phi)]_{ji}=-[R_{ij}(\phi)]_{ij}=\sin(\phi)$.

% ===============================================
\section{Coulomb integrals in anisotropic \KP}
\label{app:kpX}

To calculate the intravalley exchange anisotropy in \KP\ we use the anisotropic model and formalism proposed in Ref.~\cite{Avdeev2023g}.
We consider spherical QDs with infinite boundary conditions.
The anisotropic \KP\ Hamiltonian is expanded into three terms
\begin{equation}
  \label{eq:Haniso_kp}
  \hat{H}^{\text{aniso}} = \hat{H}^{\text{iso}} + \delta \hat{H}_P + \delta \hat{H}_{\alpha}\,,
\end{equation}
where $\hat{H}^{\text{iso}}$ is the isotropic part of the Hamiltonian, $\delta \hat{H}_P$ accounts for the anisotropy of the interband momentum matrix elements and $\delta \hat{H}_{\alpha}$ accounts for the anisotropic far-band contributions to the effective masses, see Ref.~\cite{Avdeev2023g} for details.

Eigenstates of $\hat{H}^{\text{aniso}}$ are expanded over finite basis of the solutions of the isotropic Hamiltonian $\hat{H}^{\text{iso}}$
\begin{equation}
  \label{eq:aniso_ansatz}
  \Psi_s^{\text{aniso}} = \sum_{p=1}^N C_p^s \Psi_p^{\text{iso}}
\end{equation}
and the coefficients $C$ are found via diagonalization of the full $N\times N$ matrix $\hat{H}_{qp}^{\text{aniso}}$.
Here indices $q,p$ enumerate both valence and conduction band states.
The isotropic wave functions~\cite{Avdeev2023g}
\begin{equation}
  \label{eq:FpnFz}
  \Psi^{\text{iso}} \equiv 
  \ket{F,p,n,F_z} =
  \begin{pmatrix}
    f_{F-\frac{p}2,p}\left(\frac{r}{R}\right)
    \hat \Omega_{F,F_z}^{F-\frac{p}2} \\
    \rmi p g_{F+\frac{p}2,p}\left(\frac{r}{R}\right)
    \hat \Omega_{F,F_z}^{F+\frac{p}2}
  \end{pmatrix}
\end{equation}
are bispinors characterized by four quantum numbers: total angular momentum $F$, its projection $F_z$, $p=\pm1$ and main quantum number $n$ which enumerates the roots of dispersion equation.
$\hat \Omega$ are spherical spinors.
To catch the exchange anisotropy in our calculations we limit ourselves with two ground confinement levels per band for each $F=1/2$ and $3/2$ ($N=24$ states total).

Using the ansatz \eqref{eq:aniso_ansatz} the exchange matrix elements~\cite{Avdeev2020} become sums of integrals on $\Psi_p^{\text{iso}}\equiv\ket{p}$
\begin{equation}
  \label{eq:HX_pqrs}
  H_{ia,jb}^X = \sum_{pqrs=1}^N C_p^{j*}C_q^{a*}C_r^bC_s^i \left<pq\middle|rs\right>
\end{equation}
which are evaluated numerically.
The total number of integrals scales as $N^4$, which is $331776$ for $N=24$.

To reduce the number of integrals to compute we utilize the angular selection rules and employ the index permutation symmetry.
For a general four-index Coulomb integral $\brakt{pq}{rs}$ there is particle permutation symmetry $\brakt{pq}{rs}=\brakt{qp}{sr}$, and $\brakt{pq}{rs}=\brakt{rq}{ps}=\brakt{ps}{rq}$ if the matrix elements are real.
In our case the matrix elements are real therefore one has to compute only the matrix elements which satisfy one of the following inequalities:
\begin{equation}
  \left[
  \begin{array}{l}
    p\le r\le q\le s\,,\\
    p <  r\le s <  q\,,\\
    p\le q <  r\le s\,,\\
    p <  q\le s <  r\,,\\
    p\le s <  r\le q\,,\\
    p <  s <  q <  r\,.\\
  \end{array}
  \right.
\end{equation}
These inequalities reduce the number double integrals to be evaluated for Eq.~\eqref{eq:pqrs_L} up to four times.
Combined with angular symmetry the total number of different integrals reduces to $298$.

The explicit form of the exchange integral~\eqref{eq:HX_pqrs} is
\begin{equation}
  \label{eq:pqrs_r}
  \left<pq|rs\right> = \frac{e^2}{\varepsilon_{\infty}}\int \frac{d\mathbf{x}_1 d\mathbf{x}_2}{|\bm r_1-\bm r_2|} {\Psi}_p^{*}(\mathbf{x}_1){\Psi}_q^{*}(\mathbf{x}_2) {\Psi}_r(\mathbf{x}_1) {\Psi}_s(\mathbf{x}_2),
\end{equation}
where $\int d\mathbf{x}=\int d\bm r\sum_{\sigma}$ is the spatial integral and the sum over the bispinor indices $\sigma$.
Using the Fourier decomposition of $1/r$ the Coulomb integral \eqref{eq:pqrs_r} reduces to
\begin{equation}
  \label{eq:pqrs_k}
  \left<pq|rs\right> = \int \frac{d\bm k}{2\pi^2k^2}  I_{pr}(\bm k) I_{qs}(-\bm k),
\end{equation}
where
\begin{equation}
  \label{eq:I_uv_k}
  I_{uv}(\bm k) = \int d\bm r \rme^{\rmi\bm k\bm r} 
  \left(  f_{u}f_{v}\hat \Omega_{u,-}^{\dag}\hat \Omega_{v,-} +
  p_u p_v g_{u}g_{v}\hat \Omega_{u,+}^{\dag}\hat \Omega_{v,+}
  \right).
\end{equation}
Here $f(r), g(r)$ are smooth envelopes, $\hat\Omega_{\pm}\equiv \hat\Omega_{F,F_z}^{F\pm\frac{p}2}$ are spherical spinors and indices $p,q,r,s$ and $u,v$ denote all four quantum numbers of the isotropic wave functions~\eqref{eq:FpnFz}.
By introducing \cite{Varshalovich}
\begin{equation}
  \hat \Omega_{J_1M_1}^{L_1\:\dag}\hat \Omega_{J_2M_2}^{L_2} = 
  \sum_{L=|L_1-L_2|}^{L_1+L_2} W_{L,M_1,M_2} Y_{L,-M_1+M_2},
\end{equation}
observing (numerically) that
\begin{equation}
  \label{eq:Omega_observation}
  \hat \Omega_{J_1M_1}^{J_1+\frac{p_1}2\:\dag}\hat \Omega_{J_2M_2}^{J_2+\frac{p_2}2} = 
  \hat \Omega_{J_1M_1}^{J_1-\frac{p_1}2\:\dag}\hat \Omega_{J_2M_2}^{J_2-\frac{p_2}2},
\end{equation}
and using the plane wave expansion
\begin{equation}
  \label{eq:eikr_series}
  \rme^{\rmi\bm k\bm r} = 4\pi \sum_{l=0}^{\infty}\sum_{m=-l}^l \rmi^l j_l(kr) Y_{l,m}^*(o_{\bm k}) Y_{l,m}(o_{\bm r})
\end{equation}
we further simplify $I_{uv}$ to
\begin{equation}
  \label{eq:I_uv_sum_L}
  I_{uv}(\bm k) = \sum_{L=|L_u-L_v|}^{L_u+L_v}4\pi\rmi^L W_{L,M_u,M_v} Y_{L,-M_u+M_v}(o_{\bm k}) J_{uv}^L(k)\,,
\end{equation}
where $L_u=F_u\pm\frac{p_u}2, L_v=F_v\pm\frac{p_v}2$ and
\begin{equation}
  J_{uv}^L(k) = \int_0^R r^2dr j_L(kr) \left[f_{u}f_{v}+p_up_vg_{u}g_{v}\right].
\end{equation}

Finally we reduce the Coulomb integral \eqref{eq:pqrs_k} to the sum of double integrals
\begin{multline}
  \label{eq:pqrs_L}
  \brakt{pq}{rs}=8\,\delta_{-M_p+M_r,M_q-M_s}(-1)^{-M_q+M_s}\\
  \sum_{L=L_{\min}}^{L_{\max}} W_{L,M_p,M_r}W_{L,M_q,M_s}
  \int_0^{\infty} dk J_{pr}^L(k) J_{qs}^L(k),
\end{multline}
where the $L$ limits are
\begin{equation}
  \begin{array}{l}
    L_{\min}=\max\left(|L_p-L_r|,|L_q-L_s|\right),\\
    L_{\max}=\min\left(L_p+L_r,L_q+L_s\right).
  \end{array}
\end{equation}
Now the numerical integration is straightforward.

\end{appendix}

\bibliography{uv}
%\bibliography{submission.bbl}

\end{document}